# From Fear to Self-Expression: The Contextual Nature of Physics Students' Motivations

Ben Van Dusen and Valerie Otero

*School of Education, University of Colorado, Boulder, 80309, USA*

**Abstract.** This study utilizes a sociocultural interpretation of Self-Determination Theory to better understand the role that learning contexts play in generating student motivation, engagement, and identity. By drawing on previous motivation research we develop a model that describes how a student's sense of belonging in a social setting can transform their goals and experiences. We use the extremes of fear and integrity to model a student's motivation to engage in activities. A student's sense of connection and belonging (or not) in a social setting drives whether she feels integrated with or alienated from her environment. Our model is based on three studies and suggests that a sense of belonging emerges through the alignment of goals and practices of the individual and an activity. This model is applied to two examples to illustrate how social connection or isolation can be exhibited in a physics classroom setting. We conclude by discussing the role of the teacher in designing classroom environments that support students engaging.

**Keywords:** Motivation, identity, fear, integrity, self-determination theory, competence, autonomy, and relatedness
**PACS:** 01.30.Cc, 01.40.ek, 01.40.Fk, 01.40.Ha

## INTRODUCTION

The authors' work is driven by a desire to create physics learning environments that are motivating and engaging for students. All too often physics classes leave students feeling uninterested and disconnected from the practices of science [1]. This paper outlines our process of creating a sociocultural model of motivation. We conclude by applying our model to two classroom examples and by providing recommendations for the design of physics learning environments.

## LITERATURE REVIEW

In an investigation of high school physics students' experiences, Ross and Otero [1] and Ross [2] describe two competing narratives that high school students used to describe their past and present high school experiences in science. One narrative can be characterized as that of *fear of failure* and *preservation of self-esteem*. Students used terms such as "afraid, scared, judged, stupid, boring, gullible" and "looked down upon" when they talked about their experiences in science class. In contrast, when describing their experiences in a supportive science classroom environment, students used terms consistent with *integrity* and *self-expression* such as "comfortable, interested, evidence, it's okay, legit, help each other, share," and "we have the answers." In this paper we further explore these two extremes in efforts of developing a theoretical perspective on motivation and identity. We will argue in this paper that the differences between these two extremes have to do with the extent to which students feel connected and integrated with their classroom science environments.

Ames [3] outlined a set of students' classroom motivational dispositions similar to the findings of Ross [2], in terms of students' *goals*. She argued that students either engage in classroom activities through *performance goals* (externally motivated goals associated with one's self worth) or through *mastery goals* (internally motivated goals of self-expression and inventiveness). Ames used the idea of *performance* versus *mastery goals* to argue that it is the characteristics of classrooms, not only characteristics of students, which increase the likelihood that students will perform either to protect their self-worth or out of self-expression. In our own terms, Ames [3] showed that outcomes were largely attributable to whether the classroom context engendered an orientation toward *fear and self-preservation* or whether these contexts engendered an orientation toward *integrity and self-expression*—these are phrases that capture the extremes of engagement. Ames emphasizes that the constructs of motivation and goals are determined by the nature of the context and how people see themselves in relation to that context.

Classic Self-Determination Theory (SDT) considers identity and motivation as properties of an individual in relation to social environments that shift over time and contexts [4], [5]. According to this perspective, as an individual engages within a particular context, she may become integrated (or alienated) from the activity/context and from the members of the community represented by the activity. The process of integration can shift one's motivation to engage in an activity from externally driven to being internally driven.

According to Deci and Ryan [4], internalization can be thought of as the assimilation of behaviors that were once external to the self. SDT states that internalization occurs as an activity fulfills an individual's *basic psychological needs* [4], [5]. These needs are defined as the, "conditions that are essential to an entity's growth and integrity" [5, p. 410]. In the case of the human psyche, the conditions for growth (or basic psychological needs) are: a sense of *competence* (that one can be successful), *autonomy* (that one has choice in their actions), and *relatedness* (connection to peers) [5]. When people engage in an activity that provides them the experiences of competence, autonomy, and relatedness (as contrasted with overwhelming challenges, excessive controls, and relational insecurity) they will be more likely to internalize the activity and engage in it in the future.

## MODEL DEVELOPMENT

SDT provides a framework from which to understand student motivation, however it fails to capture the dynamic nature of a person's motivation in context. For example, a student may demonstrate significant motivation when doing physics in her normal classroom environment but be completely unmotivated when moved to a physics class in a new school. By blending the work of several researchers we create a sociocultural model of student motivation and identity that incorporates and explains their contextual natures.

### *Motivation: A Sociocultural Interpretation*

Integrating the work of Ames [3], Ross [2], and Ross & Otero [1] with SDT [4] we developed the "motivation spectrum" shown in figure 1. We stress that the term motivation is not being used here with the traditional attribution to internal characteristics of an individual. Instead, "motivation," "integrity," and "identity" are used in sociocultural ways—emphasizing the role of community and context in determining the extent to which one is integrated with them (or has integrity). "Integrity" is thus defined as the extent to which the goals and practices of an activity or community have become subjective (a part of the individual) rather than objective (external to the individual), and constitutes motivation.

Differentiated motivation (shown on the left side of figure 1) emerges when an individual does not feel connected to (or integrated with) her social environment (peers and community practices); her goals and practices are out of sync with those of the activity, and the activity feels *done to her*. In such a case, a person's sense of competence and autonomy are *socially-normed*, or normed against her interpretation of the practices of a broader community. We refer to the motivation in these environments as being *differentiated*, because the students' emergent motivations and goals are distinguishable from those of the community.

On the right end of the figure, when a person feels connected to, and in control of, her social environment her actions are driven by integrity. Integrated motivation emerges when an activity becomes subjective and the individual can barely tell the difference between her goals and the goals of the community. In such a case, a person's sense of competence is *self-normed,* as distinguished from *socially-normed*. When an activity is subjective and integrated with an individual and her surrounding community her identity is entangled in the activity.

We have defined *fear and self-preservation* versus *integrity and self-expression* as representing the extremes of the motivation spectrum, however we assume that individuals' motivations in various contexts typically fall somewhere within the spectrum, and evolve through engagement in activities. In some cases, such as those on the far left of the spectrum, the individual drops out completely or maintains the same level of detachment for a long time. In many cases however, identities reflexively evolve as an individual interacts with and modifies behaviors as a result of feedback from the community.

Our theoretical model highlights the critical nature of sociocultural factors, specifically how one relates to peers and cultural practices in the emergence of identity. For example, when an individual expresses a physics identity, she cannot tell the difference between her own goals and practices and those of physics—she is integrated with physics.

### *Contextualizing Basic Psychological Needs*

As described earlier, SDT is founded on the idea of Basic Psychological Needs (competence, autonomy, and relatedness). Our review of the literature and exploratory factor analysis [6] led us to investigate *how* competence, autonomy, and relatedness were expressed, rather than *whether* they were expressed (as SDT studies have traditionally done [7]).

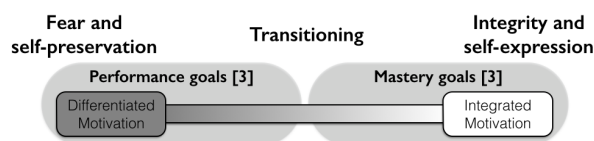

**Figure 1.** A spectrum of motivations.

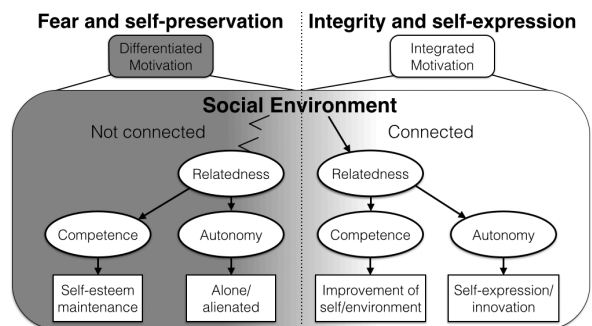

**Figure 2.** A model of contextual expression.

Figure 2, shows how fear versus integrity are largely driven by feelings of isolation versus social connectedness. It should be noted that in this model, one does not have to be in the physical presence of a social environment to feel connected to it. The constructs *competence* and *autonomy* are expressed differently depending upon an individual's level of social connectedness. A context in which an individual feels isolated (left side of figure) will lead to competence being experienced as efforts to preserve self-esteem and autonomy being experienced as alienation. A context in which an individual feels highly integrated (right side of figure) will lead to competence being experienced through efforts to improve the self and environment and autonomy being experienced as self-expression and innovation. For example, a student who does not feel socially connected to other students, the teacher, or the goals and practices of physics is afraid to express his ideas (competence), unlikely to take risks in problem solving, and may criticize the teacher and the content in an effort to not look like the one who is responsible for not fitting in (autonomy). A student who is friends with her classmates and feels connected to the practices of her physics class is more willing to take risks and try out new ideas (autonomy) with the hope of better understanding physics (competence).

Figure 2 illustrates that both competence and autonomy emerge in both types of settings, however integration within the social environment frames *how* these constructs are experienced and expressed. In order to have positive nurturing feelings of one's competence and autonomy a student must feel some connection to his social environment.

## CLASSROOM APPLICATIONS

Our model suggests that contexts can be altered to facilitate a student's sense of belonging by creating activities that are in better alignment with his/her goals and practices. In this section we apply our model to two examples from an AP physics class documented in a previous study [6]. In each example we provide a brief description of the context, the student's behaviors, and an analysis of how the interactions could be interpreted. While these two examples are abbreviated excerpts we use them in this study to illustrate potential interpretations of students' behaviors.

*Example #1:* The first example comes from AP physics student's (Lih-Hann) creation of twenty physics problem solutions in his notebook. After completing the notebook solutions at home, student groups were randomly assigned problems to share with the class using whiteboards. Lih-Hann's twenty solutions were completed in approximately half of a page in his notebook. The solutions had only 49% of the steps that the teacher had asked students to show. The final answers were correct on 58% of the problems. The percentages of complete and correct solutions are in line with the rest of the class's performance on the problem set [6].

One interpretation of Lih-Hann's actions is that they were characteristic of a performance goal (finishing the assignment as quickly as possible), that his work was socially normed, and he was doing it for someone else (the teacher). This is indicative of differentiated motivation where the goals of the student are distinct from those of the activity.

This interpretation is supported by students creating their notebooks solutions in an isolated context that offered minimal opportunities for social interaction or engagement in peer cultural practices. Students created solutions separately from their peers, the solutions were usually only seen by the students who created them and the teacher, and the expectations for their final product were based off of the AP grading guidelines. While these features may be common to many physics classes, they are not particularly conducive to fostering a sense of *connection* to students. However, as the next example will illustrate, a small change the context can provide rich opportunities for students to create community connections.

*Example #2:* The second example is Lih-Hann's creation of a screencasting solution. Screencasting apps create movies by merging sounds from the iPad's external microphone and video from the iPad's screen. The screencasts were not formally graded by the teacher but were uploaded to the course website to help fellow students learn the material. There are several features of Lih-Hann's screencast that indicate that he might have been engaging more out of integration. When working out his solution (4 min) Lih-Hann used his normal voice but when reading the problem he used a humorous high-pitch voice. Once Lih-Hann reached his final answer he shared that it did not match the answer in the back of the book. Lih-Hann discussed the differences between the two answers and how they should be interpreted. During this time, Lih-Hann expended significant effort trying to determine whether his answer was correct and how others might use his solution. It appears that Lih-Hann does not feel intimidated by the fact that he had the "wrong answer." He said:

> It equals, 6,078N. But I actually checked to see what answer we're supposed to get and it says $6.8 \times 10^3$N. So I don't know about *that*. You just saw what I did here, so it's possible I messed something up. But all I know is that these are the numbers I got, I followed the equations, and this is the answer that I came up with here: 6,078N. But according to this, it's supposed to be $6.8 \times 10^3$N. So if you see a mistake in this process go ahead and change it to try to get this answer here, *I guess*. Like I said, I don't see why this is the answer. But try to get $6.8 \times 10^3$N using this method correcting wherever I made my mistake. And if my answer is the right answer, that's how you get it. But it's not though. So aim for this, but hopefully this process will help everyone. (Lih-Hann, 3/7/13)

Lih-Hann's ability to express his disagreement with the physics textbook solution can be interpreted as evidence of integrity and self-expression. He took a chance by expressing to the whole class his open disagreement with the textbook authors. Instead of choosing another problem to solve, Lih-Hann invited the class to engage in his struggle with the difference between his answer and the answer in the text. Lih-Hann's integration with his social environment can be seen in the way that he directly addresses his audience as if he was in an active conversation with them. Lih-Hann's use of a funny voice when reading the question is indicative of him experiencing autonomy through innovation and self-expression. The significant amount of time and effort that Lih-Hann put into creating and sharing his solution, even though it was in disagreement with the textbook, is indicative of experiencing competence as attempts to improvement his environment and himself.

In this example, we illustrate how a student expressed integrity and self-expression in a physics-centric activity. In this classroom there were several features that may have contributed to students feelings of connection and belonging to their learning environment: (1) The teacher required that each student "publicly" share their homework solutions as screencasts on the class website, (2) these screencasts provided many opportunities for creativity, self-expression, and innovation, (3) the teacher required that the students assess one another on the basis of a class-generated rubric, (4) the rubric was established (with the teacher's guidance) and agreed upon by the students in the class. This environment was shown in Van Dusen [6] to increase student performance and motivation through improved student connections to the class, other students, and physics [8].

The four classroom features listed above, along with the two examples, suggest that *learning environments* can be modified in order to promote student motivation. Educators have classically thought of motivation as an intrinsic property of the individual, something that a teacher could do nothing about. Our research demonstrates that although the individual is a critical player in motivation, so too is the context that is designed, in part, by the teacher. Student motivation can be altered by creating a context in which competence and autonomy are more likely to be *expressed* as improvement of the self/environment and self-expression/innovation rather than as self-esteem maintenance and alienation. This is encouraging because it sets the stage for teachers to have input on factors that can lead students to engage or disengage.

## CONCLUSION

Our empirical [6] and theoretical work has led us to create a contextualized model of student motivation and identity. This model highlights the importance of students having connections to their social environments. These types of connections are created when the goals and practices of the students and the activities are in alignment and students engage out of integrity and self-expression.

A student's disinterest in his physics class may have little to do with his lack of abilities or his interests in physics. A confrontational or sullen student is often acting out of fear and self-preservation due to an all too common lack of connection to his social environment. Instructors should consider what it is about the environment that is making students feel isolated (or not) from an activity.

There are no surefire tricks to creating engaging physics learning environments but it is only when students feel accepted and valued that they can begin to take-up and embody the goals and practices of the physics community.

## REFERENCES

[1] M. Ross and V. Otero, *Phys. Educ. Res. Conf. Proc.*, (2012)
[2] M. Ross, *Dissertation*, CU Boulder, (2013)
[3] C. Ames, *J. Educ. Psychol.*, **84,** 3 (1992)
[4] R. Ryan and E. L. Deci, *Am. Psychol.*, **55,** 1 (2000)
[5] R. Ryan, *J. Pers.*, **63,** 3 (1995)
[6] B. Van Dusen, *Dissertation,* CU Boulder, (2014)
[7] www.selfdeterminationtheory.com (6/3/2014)
[8] S. Dykstra, B. Van Dusen, & V. Otero, *Phys. Educ. Res. Conf. Proc.*, (2014)